# Astrophysical S-factor of $^4$He$^{12}$C radiative capture at low energies

**S.B. Dubovichenko, A.V. Dzhazairov-Kakhramanov**

V.G. Fessenkov's Astrophysical Institute, 050020, Almaty, Kazakhstan

E-mail: sergey@dubovichenko.ru

**Abstract.** The possibility to describe the astrophysical S-factor of the $^4$He$^{12}$C radiative capture is considered in the potential cluster model at the energy range 0.1-4.0 MeV. It is shown that the approach used, which takes into account E2 transitions only, gives a good description of the new experimental data for adjusted parameters of potentials and leads to the value $S(300) = 16.0 \cdot$keV b.

## 1. Introduction

The process of radiative capture $^{12}$C($^4$He,$\gamma$)$^{16}$O along with the triple helium capture (Salpeter process) takes place in the thermonuclear reaction cycle of stars on their hot stage of evolution, when the temperature in the star equals hundred of millions degrees Kelvin [1]. Such high temperatures give enough energy to interacting particles for increasing the probability of passing through the Coulomb barrier, and it means that the output of such a reaction is obviously increased. This reaction leads to the formation of the stable $^{16}$O nucleus, which is the transitional link in the process of the formation of heavier elements, for example with the help of reactions $^{16}$O($^4$He,$\gamma$)$^{20}$Ne and $^{20}$Ne($^4$He,$\gamma$)$^{24}$Mg etc. However, for a long time there has been a high uncertainty in the exact determination of the $^{12}$C($^4$He,$\gamma$)$^{16}$O reaction rate, but the new recent experimental data at the energy range 1.9-4.9 MeV [2], which were measured with a good accuracy, seem to clear most of uncertainties.

Earlier we have shown the possibility of describing the astrophysical S-factor of lightest nuclei on the basis of the potential cluster model with forbidden states (FS). This model takes into account the supermultiplet symmetry of the wave function (WF) of the cluster relative motion with the splitting of orbital states according to Young's schemes [3]. The used classification of orbital states allows us to analyze the structure of inter-cluster interactions, define allowed (AS) and forbidden states in the interaction potential, and thus, the number of radial WF nodes [4].

In this work we consider the astrophysical S-factor of the $^{12}$C($^4$He,$\gamma$)$^{16}$O radiative capture on the basis of the potential cluster model with forbidden states to different states of the $^{16}$O nucleus at the energy range from 0.1 up to 4.0 MeV, and for comparison of the results of our calculations we mainly use the new experimental data form work [2].

## 2. Interaction potentials and phase shifts of scattering

First note that it is impossible to carry out the accurate classification of FS and AS for the 16-particle system, except for the S state of the $^4$H$^{12}$C system, because of the absence of tables of direct interior product of Young's schemes for A>8. Therefore, the qualitative information about the number of FS in the defined partial wave is used for the construction potentials of scattering and bound states (BS). That is, the potential of each next partial wave usually has one forbidden state less, approximately the same tendency we have in lower cluster systems with nuclei whose masses are less than or equal to eight [5].

More accurate $^4$H$^{12}$C interaction potentials for calculations were obtained in our early work [6], which now can be used for calculations of astrophysical S-factors of the $^{12}$C($^4$He,$\gamma$)$^{16}$O capture reaction, practically at any low energies. The adjustment of interaction parameters for the ground bound states is explained by the fact that during transitions from the continuum states at energies about 1-100 keV to different bound states of the $^{16}$O nucleus the maximal accurate reproduction of the nucleus bound energy is required.

For these calculations, we have rewritten our computer program, based on the finite-difference method (FDM), from TurboBasic language to Fortran-90. It allowed us to raise essentially the accuracy of all calculations, including calculations of the bound energy of the nucleus in the two-particle channel. Now the absolute accuracy of the determination of the energy of bound levels of the $^4$He$^{12}$C system really equals $10^{-6}$ MeV. The accuracy of finding the determinant's radical is about $10^{-15}$ and Wronskians of Coulomb functions are about $10^{-15}$-$10^{-20}$ [7].

Let's examine the classification of orbital states in the $^4$He$^{12}$C system, where spin S and isospin T are equal to zero. Such classification allows us to determine the total number of FS in the S interaction potential. Possible orbital Young schemes of the $^4$He$^{12}$C system are defined by Littlewood's theorem [8] and in this case it gives {444} × {4} = {844} + {754} + {7441} + {664} + {655} + {6442} + {6541} + {5551} + {5542} + {5443} + {4444} [6]. Schemes {4} and {444} correspond to $^4$He and $^{12}$C nuclei in the ground state accordingly. In accordance with known principles [8] it is possible to conclude that only the {4444} scheme will be allowed and other configurations are forbidden. Particularly, all possible configurations with the number of cells in the first raw exceeding four can not be realized, because S-shell cannot contain more than four nuclei.

Using Elliott's rule [8] it is possible to define orbital moments corresponding to various Young's schemes. Then, we obtain that the ground state (GS) of the $^{16}$O nucleus with L=0 momentum may be realized for next orbital schemes {4444}, {5551}, {664}, {844} and {6442}. This result can be used for determining the number of bound forbidden states in the potential of the ground state. As far as only {4444} symmetry is allowed in the ground state, the other schemes will be forbidden and there must be four forbidden bound states and one allowed bound energy state in the $^4$He$^{12}$C system in the $^{16}$O nucleus [6].

For description of elastic scattering processes of $^4$He and $^{12}$C nuclei and bound states of these clusters the potential of inter-cluster interactions is represented as

$$V(r) = -V_0 \exp(-\alpha r^2)$$

with a spherical Coulomb interaction where $R_c = 3.55$ fm. The potential of interactions was constructed so that it would correctly describe such characteristics as the bound energy, the Coulomb radius, the Coulomb formfactor at low transmitted momenta, the electromagnetic transition probability between bound states and the partial phase shift of the elastic scattering. The interaction potential of the 1S ground state of the $^{16}$O nucleus in the $^4$He$^{12}$C channel and other potentials for bound states meeting the abovementioned requirements were found in work [6].

Here we have adjusted BS potential parameters for more accurate description of the bound energies. Particularly, for the 1S ground state of the $^{16}$O nucleus we have received:

$$V_{1S} = 256.845472 \text{ MeV}, \quad \alpha = 0.189 \text{ fm}^{-2}. \tag{1}$$

This potential has forbidden bound states at energies: -37.56; -80.80; -134.46; -197.25 MeV, in full accordance with the FS and AS classification which was done above. The bound energy was found by finite-difference method [7] to be equal to -7.161950 MeV while the experimental value equals - 7.16195 MeV [9], charge radius equals 2.705 fm for $^4$He radius: 1.671(14) fm [10] and $^{12}$C radius: 2.4829(19) fm [11]. The experimental value of the $^{16}$O nucleus charge radius equals 2.710(15) fm [9].

More accurate interaction parameters: $V_0 = 97.7285$ MeV, $\alpha = 0.111$ fm$^{-2}$ with the same Coulomb radius were obtained for the potential of 2S level with the experimental bound energy -1.113 MeV [9]. The potential accurately reproduces the channel energy and has bound forbidden states at three values of energy: -16.0, -38.2 and -66.2 MeV, and leads to the value of root-mean-square radius of 2.97 fm. It is possible to reconstruct the potential with four values of bound forbidden states, too. The parameters of this potential equal $V_0 = 143.1092$ MeV, with the same $\alpha$ and Coulomb radius and lead to FS at -16.9, -40.5, -70.6 and -106.1 MeV, but the Coulomb radius equals 3.07 fm.

During the adjustment of parameters of the bound 1P state the following values were obtained: $V_0 = 104.11325$ MeV, $\alpha = 0.16$ fm$^{-2}$, with the same Coulomb radius. This potential has bound forbidden states at energies -19.2 and -48.6 MeV and the allowed state at the energy -0.0450 MeV, in sympathy with data from [9]. It is possible to suggest the potential with three FS and depth parameter equals $V_0 = 161.2665$ MeV, $\alpha = 0.111$ fm$^{-2}$ with the same Coulomb radius. It is accurately reproduce the energy of the bound state and has FS at energies: -20.4, -52.0, -92.5 MeV.

Following the notions outlined at the beginning of the paper we can decide that second variants of 2S and 1P potentials having four and three FS respectively, which is equal to the number of FS for S and P scattering potentials, are more correct and we will use them in further calculations.

For parameters of the 1D potential the following values were obtained: $V_0 = 90.3803$ MeV, $\alpha = 0.1$ fm$^{-2}$, with the same Coulomb radius. It leads to the value of the bound energy level equal to -0.2450 MeV, in full concordance with data from [9] and has two forbidden states at -14.0 and -34.3 MeV.

Following parameters are obtained for the 1F state of the $^{16}$O nucleus: $V_0 = 191.4447$ MeV и $\alpha = 0.277$ fm$^{-2}$. The potential gives the energy of the bound state -1.0320 MeV, which is in conformity with data from [9] and has one forbidden state at the energy -38.3 MeV.

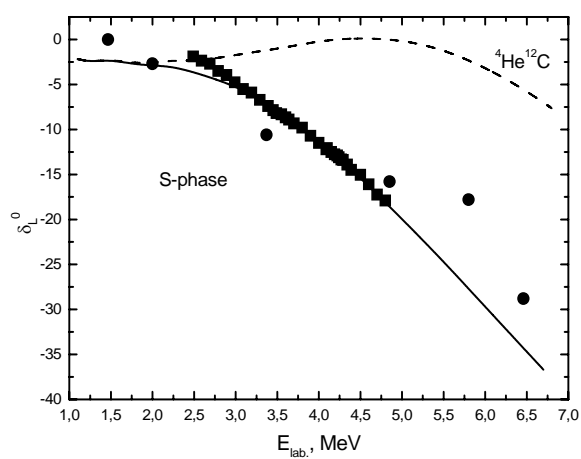

**Figure 1** - S-phase shift of the elastic $^4$He$^{12}$C scattering. Blocks - data from work [13]. Points - our results [14] obtained on the basis of experiment [12]. Curves - results of calculations with obtained potentials.

The potential of the ground 1S state doesn't lead to the correct S-phase shift of scattering, as it is shown in figure 1 by the dashed curve. For the correct description of the S phase shift obtained from the phase shift analysis [13,14] we ought to change the depth of potential and take it to be: $V_S = -155$ MeV, $\alpha = 0.189$ fm$^{-2}$ with $R_c = 3.55$ fm. The potential has four forbidden bound states at energies: -1.3, -25.1, -61.5 and -107.7 MeV and gives quite a reasonable description of S-phase shift, as it can be seen in figure 1.

New interaction potentials were obtained for P, D, F and G-waves of elastic scattering processes, which differ from bound state potentials, like in work [6]. Below are the parameters with energies of forbidden states in MeV ($R_c = 3.55$ fm):

$$V_P = 145.00 \text{ MeV}, \alpha_P = 0.160 \text{ fm}^{-2}, \quad \text{FS: } -13.6; -42.1; -79.7 \text{ MeV}; \quad (2)$$
$$V_D = 435.25 \text{ МэВ}, \alpha_D = 0.592 \text{ fm}^{-2}, \quad \text{FS: } -61.9; -167.0 \text{ MeV};$$
$$V_F = 73.40 \text{ МэВ}, \alpha_F = 0.125 \text{ fm}^{-2}, \quad \text{FS: } -7.5 \text{ MeV};$$
$$V_G = 55.55 \text{ МэВ}, \alpha_G = 0.100 \text{ fm}^{-2}, \quad \text{FS: No.}$$

Results of the phase shift calculation are shown in figures 2-5 by solid lines.

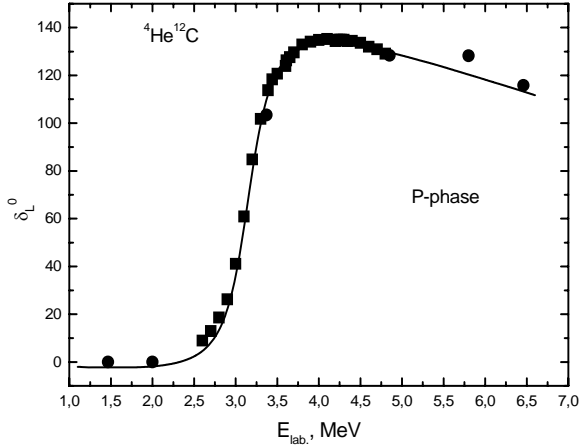
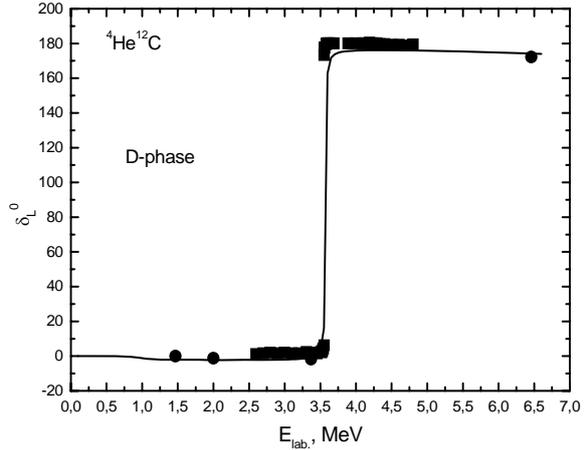

**Figure 2** - P-phase shift of the elastic $^4$He$^{12}$C scattering. The result description is the same as in figure 1.

**Figure 3** - D-phase shift of the elastic $^4$He$^{12}$C scattering. The result description is the same as in figure 1.

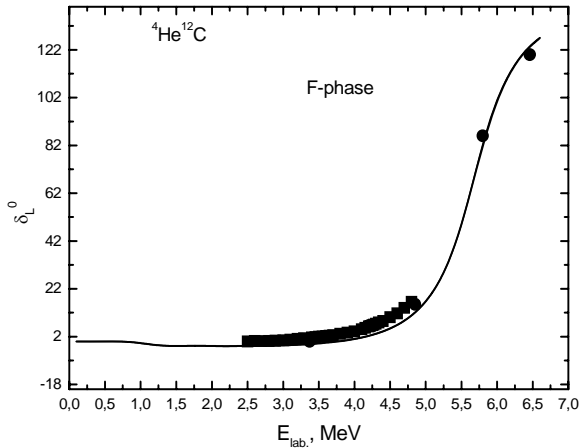
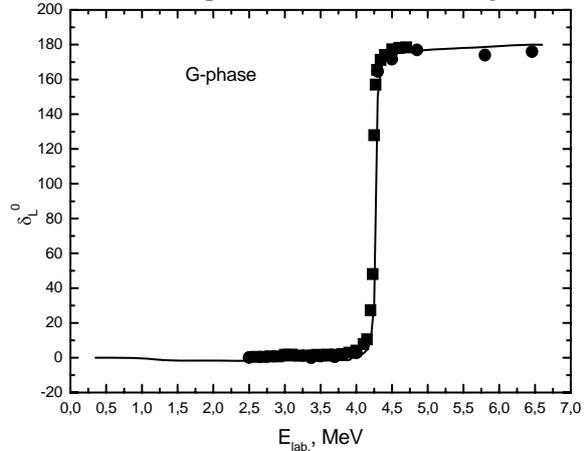

**Figure 4** - F-phase shift of the elastic $^4$He$^{12}$C scattering. The result description is the same as in figure 1.

**Figure 5** - G-phase shift of the elastic $^4$He$^{12}$C scattering. The result description is the same as in figure 1.

It seems that the obtained difference of the potential parameters describing the phase shifts of the scattering and BS characteristics can be explained by a small contribution of the considered channel into bound states of the $^{16}$O nucleus. Probably, the simple $^4$He$^{12}$C cluster model is not able to give the complete description of different characteristics of the $^{16}$O nucleus on the basis of unified potentials, as it is for lighter nuclei [5,15]. Another cause of such disagreement may be, as it was said, the absence of the accurate classification of FS and AS for the 16-particle system for all partial waves.

## 3. Astrophysical S-factor

In present S-factor calculations we use the well-known formula [16]

$$S(EJ) = \sigma(EJ)E_{cm} \exp\left(\frac{31.335 Z_1 Z_2 \sqrt{\mu}}{\sqrt{E_{cm}}}\right),$$

where $\sigma$ is the total cross-section of the radiative capture process (barn), $E_{cm}$ is the center-of-mass energy of particles (keV), $\mu$ is the reduced mass (atomic mass unit) and Z are the particle charges in elementary charge units. The numerical coefficient 31.335 was received on the basis of up-to-date values of fundamental constants, which are given in [17].

Total cross-sections $\sigma(EJ)$ of the radiative capture in the potential cluster model can be represented as (see, for example, [5] or [18])

$$\sigma_c(EJ) = \frac{8\pi K e^2}{\hbar^2 q^3} \frac{\mu}{(2S_1+1)(2S_2+1)} \frac{J+1}{J[(2J+1)!!]^2} |A_J(K) P_J(EJ) I_J|^2,$$

where for electric transitions EJ(L) we have [5]

$$P_J(EJ) = \delta_{S_i S_f}(-1)^{J_i+S+L_f+J} \sqrt{(2J+1)(2L_i+1)(2J_i+1)(2J_f+1)} (L_i 0 J 0 | L_f 0) \begin{Bmatrix} L_i & S & J_i \\ J_f & J & L_f \end{Bmatrix},$$

$$A_J(K) = K^J \mu^J \left(\frac{Z_1}{m_1^J} + (-1)^J \frac{Z_2}{m_2^J}\right), \qquad I_J = \langle \Phi_f | R^J | \Phi_i \rangle.$$

Here $\mu$ is the reduced mass; $K^J$ is the wave number of the γ-quant; q is the wave number of input channel particles; $m_1$, $m_2$, $Z_1$, $Z_2$ are particle masses and charges; $S_i = S_f = S = 0$.

The exact mass values of the particles were taken for all our calculations [17], and the $\hbar^2/m_0$ constant was taken to be 41.4686 MeV fm$^2$.

In present calculations we consider E1 and E2 processes with transitions from P and D scattering waves to the ground 1S state of the $^{16}$O nucleus, the level spectrum of the $^{16}$O nucleus is shown in figure 6. The first of them is possible due to the difference of the mass of the $^4$He nucleus, which is equal to 4.001506179127 [17], from the whole number. But the calculation results of the S-factor of this process are two or three orders less than the values of experimental data, although it has the right form due to the resonance behavior of the P phase shift of scattering [6].

**Figure 6** - The level spectrum of the $^{16}$O

The E2 transition from the D wave of scattering to the 1S state leads to the astrophysical S-factor which is shown in figure 7 by dot-dashed line. This transition describes well the experiment at the energy range 0.9 - 3.0 MeV, but does not describe it at the energy 2.46 MeV because the P phase shift of scattering has a resonance in this energy range. At energies from 2.5 to 3.0 MeV the calculated S-factor generally describes the position and the peak height caused by the resonance in the D wave of scattering at the energy 2.69 MeV, but the level width is slightly more than the experimental one. The value of the calculated S-factor of the 2$^+$ resonance is very sensitive to the depth of the potential in the D-wave. Its increase by 0.05 MeV leads to a sharper rise of the phase shift of scattering, almost without changing the position of the resonance, and reduces its value approximately three times.

Supposing that the experiment includes cross-sections with the transition to the 1P level, then it is possible to consider the E2 process from the P wave of scattering to the bound 1P state of the $^{16}$O nucleus at the energy 0.045 MeV (figure 6). Results of this calculation for the second variant of the bound 1P state potential are shown in figure 7 by a dashed line and give a good description of the resonance form at the energy 2.46 MeV. If we use the first variant of abovementioned 1P interactions with two FS, then the peak value of the 1$^-$ resonance is approximately two times less.

The calculation results of the E2 transition from the G wave to the 1D bound state are shown by the dot-dot-dashed line. They give the correct position and width of the maximum of 4$^+$ resonance, but its value turns out to be approximately two times less than the experimental data. It ought to be remarked that we are not succeeded in finding such parameters of the G wave potential, which give the correct description of the S-factor at the 4$^+$ resonance energy. If the potential of such a wave has one ($V_G$=110.7 MeV, $\alpha_G$=0.127 fm$^{-2}$, FS: -13.6 MeV) or two FS ($V_G$=222.4 MeV, $\alpha_G$=0.127 fm$^{-2}$, FS: -42.8; -14.6 MeV) then the calculated 4$^+$ resonance peak value is visibly less. The reduction of the number of FS to one ($V_D$=254.8 MeV, $\alpha_D$=0.592 fm$^{-2}$, FS: -57.0 MeV) or without FS ($V_D$=57.7833 MeV, $\alpha_D$=0.1 fm$^{-2}$) in the bound 1D state does not lead to the significant increase of the S-factor in the 4$^+$ resonance range.

Additionally cross-sections of the E4 transition from the G wave to the ground 1S state of the $^{16}$O nucleus and E1 transition from the G wave to the bound 1F state were considered. They turn out to be approximately several times less than the previous E2 process.

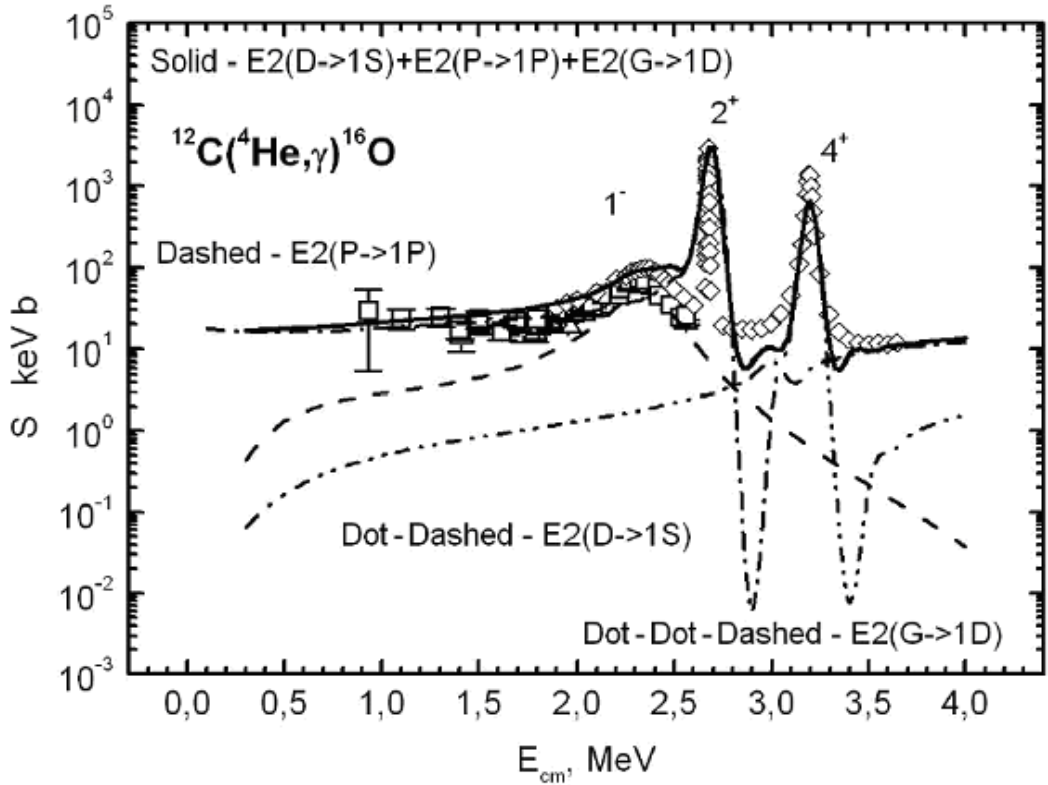

**Figure 7** – Astrophysical S-factor of $^4$He$^{12}$C radiative capture.
Open blocks denote the experimental data from [18], triangles from [19,20], open rhombs from [2].
Lines - calculations of the astrophysical S-factor for different transitions.

The sum of all three E2 transitions, which describes the experimental behavior of the astrophysical S-factor at energies from 0.9 to 4.0 MeV is shown by the solid line in figure 7.

Our calculated S-factor at the energy 300 keV caused by the E2 process with the transition from the D wave to the GS of the nucleus is equal to 16.0 keV b, and at the energy of 100 keV its value is a little higher - 17.5 keV b. However, these results are considerably less than the known data leading, for example, to $S_{E1}=79(21)$ or 82(26) keV b [21] and $S_{E2}=120(60)$ keV b [18]. Following results were obtained using the generator-coordinate method [22], taking into account different cluster configurations, $S_{E1}=160$ keV b and $S_{E2}=70$ keV b at 300 keV. Nevertheless, it doesn't follow from the available experimental data, which are shown in figure 7, that the S-factor undergoes an obvious rise and at the energy 300 keV its value is about 100 keV b.

## 4. Conclusion

Thus, the carried out qualitative analysis of the number of FS and AS in inter-cluster interactions of the $^4$He$^{12}$C system and partial potentials obtained on its basis in accordance with phase shifts of the elastic scattering and energies of bound states of the $^{16}$O nucleus give us the possibility of reasonable description of the astrophysical S-factor behavior at the energy range from 0.1 to 4.0 MeV. Such result may be considered as certain evidence in favor of the potential approach in the cluster model when the inter-cluster interactions with FS are constructed on the basis of phase shifts of the elastic scattering of clusters and each partial wave is described by its potential, for example, by the potential of the Gaussian form with specific parameters.

Such splitting of the total interaction into the partial waves allows to detail its structure and even the qualitative classification of orbital states according to Young's schemes gives the possibility to define the presence and number of forbidden states. It gives the definite depth of interaction what allows getting rid of the discrete ambiguity of the potential depth what was the case in the optical model. The form of each partial phase shift of scattering can be described correctly only for the definite width of such potential, what helps to avoid the continuous ambiguity which exists in the classical optical model too.

Certainly, all the abovementioned is correct only in case of accurate determination of experimental data for phase shifts of the elastic scattering. But, till now, phase shifts of the majority of light nuclear systems were found with considerable errors, which sometimes were equal to 20-30%. This fact complicates very much the construction of accurate potentials of the inter-cluster interaction and, in the final analysis, leads to big ambiguities in final results obtained in the potential cluster model. In this case the abovementioned information is related to some ambiguity of the D wave potential of the $^4$He$^{12}$C elastic scattering.